\documentstyle[amssymb,12pt,thmsa,a4,sw20lart]{article}

\input tcilatex

\begin{document}

\topmargin 0pt \oddsidemargin 5mm

\setcounter{page}{1}

\hspace{8cm}{} \vspace{2cm}

\begin{center}
{\Large On some generalized stopping power sum rules\bigskip }

$^{(a)}\bigskip $Hrachya B. Nersisyan\footnote{%
Corresponding author. e-mail: hrachya@irphe.am} and $^{(b)}$Amal K. Das%
\footnote{%
e-mail: akdas@is.dal.ca}

$^{(a)}${\em Division of Theoretical Physics, Institute of Radiophysics and
Electronics,}

{\em 2 Alikhanian Brothers St., Ashtarak-2, 378410, Armenia}\\[0pt]

$^{(b)}${\em Department of Physics, Dalhousie University,}

{\em Halifax, Nova Scotia B3H 3J5, Canada}
\end{center}

\centerline{{\bf{Abstract}}}

The Lindhard-Winther (LW) equipartition sum rule shows that within the
linear response theory, the stopping power of an energetic point-charge
projectile in a degenerate electron gas medium, receives equal contributions
from single-particle and collective excitations in the medium. In this paper
we show that the LW sum rule does not necessarily hold for an extended
projectile ion and for ion-clusters moving in a fully degenerate electron
gas. We have derived a generalized equipartition sum rule and some related
sum rules for this type of projectiles. We also present numerical plots for
He$^{+}$ ion and He$^{+}$ ion-clusters.

\newpage

\section{Introduction}

The stopping power (SP), which is a measure of energy loss of energetic
charged particles in a target medium, is of continuing theoretical and
experimental interest in diverse areas such as interaction of charged
particles with solids (see [1-4] for reviews) and beam-heating of fusion
plasma \cite{5}. For high-velocity projectile particles or clusters, the
energy loss may be mainly due to collective and single-particle excitations
in the target medium. It is of fundamental and practical interest to study
the extent to which the collective and single-particle excitations each
contribute to SP. This is the objective of our work reported here.

The energy loss of a high velocity projectile is formulated, originally due
to Lindhard \cite{6}, on the justifiable assumption of a weak coupling
between the particle and a target medium which is modelled by a degenerate
electron gas (DEG), through the linear response function of the DEG. The
corresponding dielectric function $\epsilon (k,\omega )$ contains
contributions from both collective excitations (plasmons) and
single-particle excitations. For a single point-ion projectile Lindhard and
Winther (LW) \cite{7} investigated the respective contributions of these two
excitations and found a sum rule which states that both these excitations
contribute equally to SP. To our knowledge this type of sum rule has not yet
been studied for an extended-charge projectile or a cluster. In this paper
we formulate a generalized stopping power sum rule, again in the linear
response approach, and present mostly analytical results for a He$^{+}$ ion
and for a diproton cluster as projectiles. We compare and contrast our
results with those of LW.

In linear response theory, the stopping power (SP) which is the energy loss
per unit length for an external projectile with a spatial charge
distribution $\rho _{{\rm ext}}({\bf r},t)=Q_{{\rm ext}}({\bf r}-{\bf V}t)$
moving with velocity ${\bf V}$ in a homogeneous isotropic medium
characterized by the dielectric function $\varepsilon (k,\omega )$, is given
by (see, e.g., \cite{3})

\begin{equation}
S=\frac{1}{2\pi ^{2}V}\int d{\bf k}\left| G({\bf k})\right| ^{2}\frac{{\bf %
k\cdot V}}{k^{2}}{\rm Im}\frac{-1}{\varepsilon (k,{\bf k\cdot V})}{\bf ,}
\end{equation}
where $G({\bf k})$ is the Fourier transform of the stationary charge $Q_{%
{\rm ext}}({\bf r})$.

Eq. (1) is applicable to any external charge distribution. In Sec. II we
discuss a dicluster of two identical He$^{+}$ ions separated by a variable
distance ${\bf R}$. The spatial distribution of bound electrons in the ions, 
$\rho (r)$, is assumed to be spherically symmetric. We use a 1s-type wave
function of the form $\psi _{1s}(r)=\left( Z^{3}/\pi a_{0}^{3}\right)
^{1/2}e^{-Zr/a_{0}}$, to describe the bound electron on each He$^{+}$ ion,
with $a_{0}=0.529$ \AA\ as the Bohr radius. $Ze$ is the charge on each of
the point-like nuclei ($Z=2$). It may be remarked that we are considering an
unscreened 1s electron. For the projectile systems under study we may write $%
G({\bf k})$ as

\begin{equation}
G({\bf k})=e\left[ Z-\rho (k)\right] \left[ 1+\exp \left( -i{\bf k\cdot R}%
\right) \right] ,
\end{equation}
where $\rho (k)$ is the Fourier transform of $\rho (r)=\left| \psi
_{1s}(r)\right| ^{2}$. $G({\bf k})$ contains ${\bf R}$ as a parameter.

For a dicluster of He$^{+}$ ions the SP of a dicluster can be written as 
\cite{3}

\begin{eqnarray}
S &=&2S_{{\rm ind}}\left( \lambda \right) +2S_{{\rm corr}}\left( \lambda
,R,\vartheta \right) \\
&=&\frac{16e^{2}k_{F}^{2}}{\pi \lambda ^{2}}\int_{0}^{\lambda
}udu\int_{0}^{\infty }{\cal Z}^{2}(z){\rm Im}\frac{-1}{\varepsilon (z,u)}%
\left[ 1+\cos \left( {\cal A}z\cos \vartheta \right) J_{0}\left( {\cal B}%
z\sin \vartheta \right) \right] zdz,  \nonumber
\end{eqnarray}
where ${\cal A}=\left( 2u/\lambda \right) k_{F}R$, and ${\cal B}=2k_{F}R%
\sqrt{1-u^{2}/\lambda ^{2}}$. $S_{{\rm ind}}\left( \lambda \right) $ and $S_{%
{\rm corr}}\left( \lambda ,R,\vartheta \right) $ stand for individual and
correlated SP, respectively. $J_{0}(x)$ is the Bessel function of first kind
and zero order and $\vartheta $ is the angle between the interionic
separation vector ${\bf R}$ and the velocity vector ${\bf V}$; $\lambda
=V/v_{F}$, $\chi ^{2}=1/\pi k_{F}a_{0}=\left( 4/9\pi ^{4}\right) ^{1/3}r_{s}$%
, $r_{s}=\left( 3/4\pi n_{0}a_{0}^{3}\right) ^{1/3}$. $n_{0}$ is the
electron gas density, $v_{F}$ and $k_{F}$ are the Fermi velocity and wave
number of the target electrons respectively. In our calculations $\chi $ (or 
$r_{s}$) serves as a measure of electron density. Here, as in Refs.
[2-4,6,7], we have introduced the following notations $z=k/2k_{F}$, $%
u=\omega /kv_{F}$. In these variables ${\cal Z}(z)=Z-\rho (z)$ and the
Fourier transform of the spatial distribution $\rho (r)$ is expressed as

\begin{equation}
\rho (z)=\frac{\alpha ^{4}}{\left( z^{2}+\alpha ^{2}\right) ^{2}},
\end{equation}
where $\alpha =\pi \chi ^{2}Z$.

In Eq. (3) the term for correlated stopping power $S_{{\rm corr}}$ vanishes
for large $R$ ($R\rightarrow \infty $) and SP is the sum of individual
stopping powers for the separate ions. For $R\rightarrow 0$ the two ions
coalesce into a single entity. Then $S_{{\rm corr}}=S_{{\rm ind}}$ and SP is
that for a total charge $2e(Z-1)$.

We will consider the interaction process of a dicluster in a fully
degenerate ($T=0$) electron gas. For this purpose we use the exact
random-phase approximation (RPA) dielectric response function due to
Lindhard \cite{6},

\begin{eqnarray}
\varepsilon (k,\omega ) &=&1+\frac{2m^{2}\omega _{p}^{2}}{\hbar ^{2}k^{2}}%
\sum_{n=1}^{N}\frac{f\left( k_{n}\right) }{N}\left\{ \frac{1}{k^{2}+2{\bf %
k\cdot k}_{n}-\left( 2m/\hbar \right) \left( \omega +i0^{+}\right) }\right.
\\
&&\left. +\frac{1}{k^{2}-2{\bf k\cdot k}_{n}+\left( 2m/\hbar \right) \left(
\omega +i0^{+}\right) }\right\} .  \nonumber
\end{eqnarray}
Here, ${\bf k}_{n}$ is the wave vector of the electron in the $n$th state.
The distribution function $f\left( k_{n}\right) $ is an even function of $%
{\bf k}_{n}$, and normalized so that $N=\sum_{n}f\left( k_{n}\right) $ is
the total number of electrons. In the case of a fully degenerate free
electron gas with Fermi energy $E_{F}=\hbar ^{2}k_{F}^{2}/2m$ the
distribution function is $f\left( k_{n}\right) =1$ for $k_{n}<k_{F}$, and $%
f\left( k_{n}\right) =0$ for $k_{n}>k_{F}$.

The summation in Eq. (5) can be analytically performed, leading to the
characteristic logarithmic expression in $\varepsilon (z,u)$ first obtained
by Lindhard \cite{6}. However, in our further consideration it is convenient
to use the form (5) for the dielectric function. This allows a wider
investigation of the analytical properties of $\varepsilon (z,u)$ in the
complex $z$ plane with applications toward deriving some useful sum rules.

\section{Equipartition sum rule}

In this section we shall discuss some important properties of the SP
integrals, which we call stopping power summation rules (SPSR). In the
literature two such sum rules have been widely considered$-$the familiar
Bethe sum rule and the Lindhard-Winther (LW) equipartition rule (see \cite{7}%
). While the Bethe sum rule concerns the integral of $1/\varepsilon (z,u)$
over $\omega $ (or $u$) for fixed $k$ (or $z$), the LW equipartition rule
concerns the integral of $1/\varepsilon (z,u)$ over $z$ for fixed $u$. The
latter summation rule states that an integral of the form

\begin{equation}
\int_{0}^{\infty }{\rm Im}\frac{-1}{\varepsilon (z,u)}zdz=\Im _{p}(u)+\Im
_{sp}(u)
\end{equation}
receives equal contributions, $\Im _{p}(u)$ and $\Im _{sp}(u)$,
respectively, from the plasma resonance (plasmons), $0<z<u-1$, and from the
region of close collisions (single-particle excitations), $u-1<z<u+1$.

The LW equipartition rule was originally formulated for a single point-like
charged projectile. Here we shall examine and generalize this equipartition
rule for extended projectiles and their diclusters. As an example we shall
consider a dicluster of two He$^{+}$ ions separated by a variable distance $%
R $. In order to deal with extended projectiles, as a generalization of Eq.
(6), we need to consider an integral proportional to that in Eq. (3),

\begin{equation}
\int_{0}^{\infty }{\cal Z}^{2}(z){\rm Im}\frac{-1}{\varepsilon (z,u)}\left[
1+\cos \left( {\cal A}z\cos \vartheta \right) J_{0}\left( {\cal B}z\sin
\vartheta \right) \right] zdz,
\end{equation}
which we decompose as $\Im _{p}(\alpha ,u)+\Im _{sp}(\alpha ,u)$, where $\Im
_{p}(\alpha ,u)$ and $\Im _{sp}(\alpha ,u)$ are the contributions for
plasmon and single-particle excitations.

In Eq.(7) the respective contributions for plasmon and single-particle
excitations can be written as the sum of individual (first term) and
correlated (second term) stopping terms

\begin{equation}
\Im _{p}(\alpha ,u)=\Im _{p}^{{\rm ind}}(\alpha ,u)+\Im _{p}^{{\rm corr}%
}(\alpha ,u),\quad \Im _{sp}(\alpha ,u)=\Im _{sp}^{{\rm ind}}(\alpha ,u)+\Im
_{sp}^{{\rm corr}}(\alpha ,u).
\end{equation}
Here again the terms for correlated stopping contribution, $\Im _{p}^{{\rm %
corr}}$ and $\Im _{sp}^{{\rm corr}}$, for $R\rightarrow 0$ tend to $\Im
_{p}^{{\rm ind}}$ and $\Im _{sp}^{{\rm ind}}$ respectively.

Our objective is to show that the LW equipartition rule, $\Im _{p}=\Im _{sp}$%
, is not necessarily satisfied for extended projectiles and their clusters.
We may therefore introduce the function

\begin{equation}
\Delta (\alpha ,u)=\Im _{sp}(\alpha ,u)-\Im _{p}(\alpha ,u),
\end{equation}
which represents the difference between single-particle and plasmon
contributions to the integral given by Eq. (7). In order to calculate the
function $\Delta (\alpha ,u)$, it is imperative to consider the integral in
the complex $z$ plane and to find the poles of $1/\varepsilon (z,u)$, i.e.
the zeros of $\varepsilon (z,u)$, for fixed $u$. It is seen from Eq. (5)
that there must be $4N+2$ zeros of $\varepsilon (z,u)$ for fixed $u$. For a
large real value of $u$, above the value $u_{0}$ which corresponds to the
minimum in the plasma resonance, all $4N+2$ zeros lie on the real $z$ axis.
Two zeros occur at $z=\pm z_{r}(u)$, and are determined by the intersection
with the plasma resonance curve. $2N$ zeros are grouped together in the
intervals $u-1<\left| z\right| <u+1$, and are responsible for the
single-particle contribution, $\Im _{sp}$, to the integral in Eq. (7). The
remaining $2N$ zeros are also grouped together near the point $z=0$, but
they lie outside of the interval $u-1<\left| z\right| <u+1$ which is
responsible for single-particle excitations. It can be seen directly from
Eq. (5) that at $z\rightarrow 0$ $\varepsilon (z,u)\rightarrow \infty $.
Therefore the contribution of these latter $2N$ zeros of $\varepsilon (z,u)$
to the integral in Eq. (7) vanishes.

It should be noted that in the RPA dielectric function (5), the variable $u$
has an infinitesimal positive imaginary part which is introduced for
causality. It is seen from Eq. (5) that a small positive imaginary part $%
i0^{+}$ being added to $u$, is equivalent to a displacement of the zeros of $%
\varepsilon (z,u)$ in the complex $z$ plane in such a way that the zeros $%
z=\pm z_{r}(u)$ at plasma resonance lie below the real $z$ axis while the
zeros in the region $u-1<\left| z\right| <u+1$, lie above this axis. From
Eq. (5) it also follows that the slope $\partial \varepsilon (z,u)/\partial
z $ is positive at the plasmon curve, $z=z_{r}(u)$ but negative at the other
zero curves, $z=z_{j}(u)$ in the $z$ plane.

With these observations we shall now establish a generalized SP
equipartition sum rule for a He$^{+}$ ion dicluster projectile, and shall
present results for two values of the orientation angle $\vartheta $ made by
the projectile velocity vector with the inter-He$^{+}$ ion separation
vector. The result for a single He$^{+}$ ion will subsequently be obtained
from the dicluster sum rule.

\subsection{He$^{+}$ ion cluster with $\protect\vartheta =0$}

To derive an expression for a non-zero $\Delta (\alpha ,u)$ and to
generalize the LW summation rule we consider the following contour integral

\begin{equation}
Q_{\pm }(\alpha ,u)=\int_{C_{1},\,D_{1}}{\cal Z}^{2}(z)\left[ 1-\frac{1}{%
\varepsilon (z,u)}\right] \left[ 1+\exp \left( \pm i{\cal A}z\right) \right]
zdz,
\end{equation}
where the contours $C_{1}$ (for $Q_{+}(\alpha ,u)$) and $D_{1}$ (for $%
Q_{-}(\alpha ,u)$) are shown in Fig. 1. These two contours contain the real $%
z$ axis ($0,+\infty $), upper (for $C_{1}$) or lower (for $D_{1}$) quarter
circles, the imaginary $z$ axis ($\pm i\infty ,0$) and infinitesimal
semicircles $C_{2}$ or $D_{2}$. The full structure of the integral in Eq.
(10), for the cases of He$^{+}$\ ion and a dicluster, is more involved than
for the case of a point-like ion, and it also contains an exponential
function. An analytical evaluation of this integral thus leads us to
consider a contour different from the one used in the LW paper (see \cite{7}%
, for details).

For large values of $|z|$, the dielectric function must behave as

\begin{equation}
\varepsilon (z,u)\rightarrow 1+\frac{\chi ^{2}}{3z^{4}},
\end{equation}
according to Eq. (5). Therefore both integrals $Q_{\pm }(\alpha ,u)$ vanish
within the upper and lower quadrants, respectively and from Eq. (10) we find

\begin{equation}
Q_{\pm }(\alpha ,u)=\left( \int_{0}^{\infty }+\int_{\pm i\infty
}^{0}+\int_{C_{2},\,D_{2}}\right) {\cal Z}^{2}(z)\left[ 1-\frac{1}{%
\varepsilon (z,u)}\right] \left[ 1+\exp \left( \pm i{\cal A}z\right) \right]
zdz.
\end{equation}

On the other hand both the integrands in $Q_{\pm }(\alpha ,u)$ are
analytical functions inside contours $C_{1}$ and $D_{1}$ containing
single-particle $N$ poles $z_{j}+i0^{+}$, $j=1,2,...,N$, (contour $C_{1}$)
or single plasmon pole, $z_{r}(u)-i0^{+}$, (contour $D_{1}$). According to
the theorem of residues, for these functions we have

\begin{equation}
Q_{+}(\alpha ,u)=-2\pi i\sum_{j}\frac{z_{j}{\cal Z}^{2}(z_{j})}{\frac{%
\partial }{\partial z}\varepsilon (z_{j},u)}\left[ 1+\exp \left( i{\cal A}%
z_{j}\right) \right] ,
\end{equation}

\begin{equation}
Q_{-}(\alpha ,u)=2\pi i\frac{z_{r}{\cal Z}^{2}(z_{r})}{\frac{\partial }{%
\partial z}\varepsilon (z_{r},u)}\left[ 1+\exp \left( i{\cal A}z_{r}\right) %
\right] .
\end{equation}
As has been mentioned earlier, $\partial \varepsilon (z_{r},u)/\partial z>0$
for plasmons and $\partial \varepsilon (z_{j},u)/\partial z<0$\ for
single-particle excitations.

Now let us take the imaginary part of both sides of Eqs. (12)-(14). We find

\begin{eqnarray}
{\cal I}_{\pm }(\alpha ,u) &=&\func{Im}\int_{0}^{\infty }{\cal Z}^{2}(z)%
\left[ 1-\frac{1}{\varepsilon (z,u)}\right] \left[ 1+\exp \left( \pm i{\cal A%
}z\right) \right] zdz \\
&=&2\left( 
\begin{array}{c}
\Im _{sp}(\alpha ,u) \\ 
\Im _{p}(\alpha ,u)
\end{array}
\right) -P_{\pm }(\alpha ,u),  \nonumber
\end{eqnarray}
where

\begin{equation}
P_{\pm }(\alpha ,u)=\func{Im}\int_{C_{2},\,D_{2}}{\cal Z}^{2}(z)\left[ 1-%
\frac{1}{\varepsilon (z,u)}\right] \left[ 1+\exp \left( \pm i{\cal A}%
z\right) \right] zdz.
\end{equation}
In Eq. (15) we have introduced

\begin{eqnarray}
\Im _{sp}(\alpha ,u) &=&-\sum_{j}\frac{\pi z_{j}{\cal Z}^{2}(z_{j})}{\frac{%
\partial }{\partial z}\varepsilon (z_{j},u)}\left[ 1+\cos \left( {\cal A}%
z_{j}\right) \right] \\
&=&\int_{u-1}^{u+1}{\cal Z}^{2}(z)\func{Im}\frac{-1}{\varepsilon (z,u)}\left[
1+\cos \left( {\cal A}z\right) \right] zdz,  \nonumber
\end{eqnarray}

\begin{eqnarray}
\Im _{p}(\alpha ,u) &=&\frac{\pi z_{r}{\cal Z}^{2}(z_{r})}{\frac{\partial }{%
\partial z}\varepsilon (z_{r},u)}\left[ 1+\cos \left( {\cal A}z_{r}\right) %
\right] \\
&=&\int_{0}^{u-1}{\cal Z}^{2}(z)\func{Im}\frac{-1}{\varepsilon (z,u)}\left[
1+\cos \left( {\cal A}z\right) \right] zdz,  \nonumber
\end{eqnarray}
for single-particle ($\Im _{sp}(\alpha ,u)$) and plasmon ($\Im _{p}(\alpha
,u)$) contributions to the integral (7), respectively. To prove that the
sums in Eqs. (17) and (18) are actually equal to the integral forms of $\Im
_{sp}(\alpha ,u)$ and $\Im _{p}(\alpha ,u)$ we use the known expression
(see, e.g., \cite{8})

\begin{equation}
\left. \func{Im}\frac{-1}{\varepsilon (z,u)}\right| _{\func{Im}\varepsilon
(z,u)\rightarrow 0^{+}}=\pi \sum_{j}\frac{\delta \left( z-z_{j}\right) }{%
\left| \frac{\partial }{\partial z}\varepsilon (z_{j},u)\right| },
\end{equation}
where $z_{j}$ are the zeros of $\varepsilon (z,u)$. Since all $N$
single-particle poles lie in the interval $u-1<z<u+1$, the integration in
Eq. (17) over $z$ results in the summation form for $\Im _{sp}(\alpha ,u)$.
However, for plasmon contribution the plasmon pole $z_{r}(u)$ lies in the
interval $0<z<u-1$, only for sufficiently high $u$, $u>u_{0}(\chi )$ \cite{7}%
. The threshold value $u_{0}(\chi )$ depends on the electron gas density.
For instance, for metallic densities $r_{s}\sim 2$ ($\chi \sim 0.5$) for
threshold value of $u$ we have $u_{0}\sim 1.4$. When $u<u_{0}(\chi )$ the
plasmon contribution term vanishes, $\Im _{p}=0$.

Next, we note that the imaginary part of the second terms in Eq. (12) has
been omitted, because the function

\begin{equation}
{\cal Z}^{2}(\pm iz)\left[ 1-\frac{1}{\varepsilon (\pm iz,u)}\right]
\end{equation}
is real. Consequently

\begin{eqnarray}
&&\func{Im}\int_{\pm i\infty }^{0}{\cal Z}^{2}(z)\left[ 1-\frac{1}{%
\varepsilon (z,u)}\right] \left[ 1+\exp \left( \pm i{\cal A}z\right) \right]
zdz \\
&=&\func{Im}\int_{0}^{\infty }{\cal Z}^{2}(\pm iz)\left[ 1-\frac{1}{%
\varepsilon (\pm iz,u)}\right] \left[ 1+\exp \left( -{\cal A}z\right) \right]
zdz=0.  \nonumber
\end{eqnarray}

Let us now consider the integrals $P_{\pm }(\alpha ,u)$. From Fig. 1 and Eq.
(16) and after evaluating the residues we find

\begin{eqnarray}
P_{+}(\alpha ,u) &=&-P_{-}(\alpha ,u) \\
&=&\frac{\pi \alpha ^{3}}{2}\left[ \left( Z-\frac{1}{16}\right) \frac{%
\partial }{\partial \alpha }+\frac{\alpha }{16}\frac{\partial ^{2}}{\partial
\alpha ^{2}}-\frac{\alpha ^{2}}{48}\frac{\partial ^{3}}{\partial \alpha ^{3}}%
\right] \Phi (\alpha ,u)  \nonumber
\end{eqnarray}
where

\begin{equation}
\Phi (\alpha ,u)=\left[ 1+\exp \left( -{\cal A}\alpha \right) \right] \left[ 
\frac{1}{\widetilde{\varepsilon }(\alpha ,u)}-1\right] ,
\end{equation}

\begin{eqnarray}
\widetilde{\varepsilon }(\alpha ,u) &=&\varepsilon (i\alpha ,u)=1+\frac{\chi
^{2}}{4\alpha ^{3}}\left\{ -2\alpha -\arctan \frac{u-1}{\alpha }+\arctan 
\frac{u+1}{\alpha }\right. \\
&&\left. +\left( u^{2}-\alpha ^{2}\right) \left( \arctan \frac{\alpha }{u+1}%
-\arctan \frac{\alpha }{u-1}\right) +u\alpha \ln \frac{(u+1)^{2}+\alpha ^{2}%
}{(u-1)^{2}+\alpha ^{2}}\right\} .  \nonumber
\end{eqnarray}

Therefore from Eqs. (15) and (22) we finally obtain

\begin{eqnarray}
\Delta _{c}^{(h)}(\alpha ,u) &=&P_{+}(\alpha ,u)+\frac{1}{2}\left[ {\cal I}%
_{+}(\alpha ,u)-{\cal I}_{-}(\alpha ,u)\right] \\
&=&P_{+}(\alpha ,u)+{\rm P}\int_{0}^{\infty }{\cal Z}^{2}(z)\func{Re}\left[
1-\frac{1}{\varepsilon (z,u)}\right] \sin \left( {\cal A}z\right) zdz. 
\nonumber
\end{eqnarray}
Note that in Eq. (25) at $z=z_{r}(u)$, where $0<z_{r}(u)<u-1$, both the real
and imaginary parts of the dielectric function vanish i.e. $\varepsilon
(z_{r},u)=0$. Therefore one needs to consider the Cauchy principal value
(denoted by ${\rm P}$) of the integral.

Eq. (25) is the generalized SPSR for a He$^{+}$ ion cluster with $\vartheta
=0$. From the general expression (25) we shall derive below some particular
SPSR for individual He$^{+}$ ion and diproton cluster with $\vartheta =0$.

\subsubsection{Individual He$^{+}$ ion}

We can derive the SPSR for an individual He$^{+}$ ion directly from Eq. (25)
if we consider the limit of ${\cal A}\rightarrow \infty $. In this limit the
exponential function in Eq. (23) eventually vanishes and the function $%
P_{+}(\alpha ,u)$ is defined by the first term in Eq. (23). The second term
in Eq. (25) in the limit of ${\cal A}\rightarrow \infty $ must behave as $%
\cos \left[ a\left( u\right) {\cal A}\right] $ as can be seen from Eqs. (17)
and (18), where $a(u)$ is some unspecified function of $u$. Therefore the
second term in Eq. (25) oscillates with an increasing ${\cal A}$ or
interionic distance $R$; the full integral of this term over $u$ is damped
although not necessarily vanishing as ${\cal A}\rightarrow \infty $.
However, when we include a small damping in the electron gas, which is
expected for any real medium, the second term in Eq. (25) vanishes as ${\cal %
A}\rightarrow \infty $. Thus for an individual He$^{+}$ ion we find:

\begin{equation}
\Delta ^{(h)}(\alpha ,u)=P_{\infty }(\alpha ,u),
\end{equation}
where $P_{\infty }(\alpha ,u)$ is the function $P_{+}(\alpha ,u)$ at ${\cal A%
}\rightarrow \infty $.

For individual protons ($\alpha \rightarrow \infty $) the right-hand side of
Eq. (26) behaves as

\begin{equation}
P_{\infty }(\alpha ,u)\simeq \frac{2\pi \chi ^{2}}{3\alpha ^{2}}\left(
Z-1\right) \rightarrow 0.
\end{equation}
Consequently, in this limit we recover the known Lindhard-Winther
equipartition rule (ER), $\Im _{sp}=\Im _{p}$.

\subsubsection{Diproton cluster with $\protect\vartheta =0$}

For a diproton cluster ($\alpha \rightarrow \infty $) the function $%
P_{+}(\alpha ,u)$ vanishes as in Eq. (27). Therefore for a cluster of
point-like particles we find

\begin{equation}
\Delta _{c}^{(p)}(\alpha ,u)={\rm P}\int_{0}^{\infty }\func{Re}\left[ 1-%
\frac{1}{\varepsilon (z,u)}\right] \sin \left( {\cal A}z\right) zdz.
\end{equation}

For individual protons ($R\rightarrow \infty $ or ${\cal A}\rightarrow
\infty $) the RHS of Eq. (28) vanishes due to a small damping in the
electron gas and again we recover the Lindhard-Winther ER.

\subsection{He$^{+}$ ion cluster with $\protect\vartheta =\protect\pi /2$}

In order to derive an analytical expression for $\Delta (\alpha ,u)$ for a He%
$^{+}$ ion dicluster with $\vartheta =\pi /2$, we use the same integration
contours $C_{1}$ and $D_{1}$ of Fig. 1, and consider the following integrals

\begin{equation}
Q_{\pm }(\alpha ,u)=\int_{C_{1},\,D_{1}}{\cal Z}^{2}(z)\left[ 1-\frac{1}{%
\varepsilon (z,u)}\right] \left[ 1+H_{0}^{(1,2)}\left( {\cal B}z\right) %
\right] zdz,
\end{equation}
where $H_{0}^{(1)}\left( z\right) $ and $H_{0}^{(2)}\left( z\right) $ are
the Hankel cylindrical functions of the first and second kind, respectively
and of zero order. We may recall that the Hankel functions are analytic
inside and on the contours $C_{1}$ (for $Q_{+}(\alpha ,u)$) and $D_{1}$ (for 
$Q_{-}(\alpha ,u)$) except at the point $z=0$ where they have a logarithmic
singularity. Moreover, the functions $H_{0}^{(1)}\left( z\right) $ and $%
H_{0}^{(2)}\left( z\right) $ vanish on the upper and lower quadrants,
respectively.

Using the theorem of residues, the functions $Q_{\pm }(\alpha ,u)$ are
evaluated as

\begin{equation}
Q_{+}(\alpha ,u)=-2\pi i\sum_{j}\frac{z_{j}{\cal Z}^{2}(z_{j})}{\frac{%
\partial }{\partial z}\varepsilon (z_{j},u)}\left[ 1+H_{0}^{(1)}\left( {\cal %
B}z_{j}\right) \right] ,
\end{equation}

\begin{equation}
Q_{-}(\alpha ,u)=2\pi i\frac{z_{r}{\cal Z}^{2}(z_{r})}{\frac{\partial }{%
\partial z}\varepsilon (z_{r},u)}\left[ 1+H_{0}^{(2)}\left( {\cal B}%
z_{r}\right) \right] .
\end{equation}

Then for $\vartheta =\pi /2$, we obtain from Eqs.(29)-(31)

\begin{equation}
\Im _{sp}(\alpha ,u)=\sum_{j}\frac{\pi z_{j}{\cal Z}^{2}(z_{j})}{\left| 
\frac{\partial }{\partial z}\varepsilon (z_{j},u)\right| }\left[
1+J_{0}\left( {\cal B}z_{j}\right) \right] ,
\end{equation}

\begin{equation}
\Im _{p}(\alpha ,u)=\frac{\pi z_{r}{\cal Z}^{2}(z_{r})}{\frac{\partial }{%
\partial z}\varepsilon (z_{r},u)}\left[ 1+J_{0}\left( {\cal B}z_{r}\right) %
\right] ,
\end{equation}
which are different from the corresponding quantities for $\vartheta =0$
given in Eqs.(17) and (18).

Using a similar procedure of calculation as in the previous section we
finally find

\begin{eqnarray}
\Delta _{c}^{(h)}(\alpha ,u) &=&P_{\infty }\left( \alpha ,u\right) +{\rm P}%
\int_{0}^{\infty }{\cal Z}^{2}(z)\func{Re}\left[ 1-\frac{1}{\varepsilon (z,u)%
}\right] Y_{0}\left( {\cal B}z\right) zdz \\
&&-\frac{2}{\pi }{\rm P}\int_{0}^{\infty }\widetilde{{\cal Z}}^{2}(z)\left[
1-\frac{1}{\widetilde{\varepsilon }(z,u)}\right] K_{0}\left( {\cal B}%
z\right) zdz,  \nonumber
\end{eqnarray}
where $\widetilde{{\cal Z}}(z)={\cal Z}(iz)$, $Y_{0}\left( z\right) $ is the
Bessel function of the second kind and zero order, $K_{0}\left( z\right) $
is the modified Bessel function of the second kind and zero order.

After some algebraic manipulation, the last term in Eq. (34) can be shown to
read:

\begin{eqnarray}
&&\frac{2}{\pi }Z^{2}\int_{0}^{\infty }\left[ 1-\frac{1}{\widetilde{%
\varepsilon }(z,u)}\right] K_{0}\left( {\cal B}z\right) zdz \\
&&+\frac{2\alpha ^{3}}{\pi }\left[ \left( Z-\frac{1}{16}\right) \frac{%
\partial }{\partial \alpha }+\frac{\alpha }{16}\frac{\partial ^{2}}{\partial
\alpha ^{2}}-\frac{\alpha ^{2}}{48}\frac{\partial ^{3}}{\partial \alpha ^{3}}%
\right] {\rm P}\int_{0}^{\infty }\frac{K_{0}\left( {\cal B}z\right) }{\alpha
^{2}-z^{2}}\left[ 1-\frac{1}{\widetilde{\varepsilon }(z,u)}\right] zdz. 
\nonumber
\end{eqnarray}

Eq. (34) is the generalized equipartition sum rule for a He$^{+}$ dicluster
with $\vartheta =\pi /2$. In the limit $\alpha \rightarrow \infty $, we
obtain the corresponding sum rule for a diproton cluster with $\vartheta
=\pi /2$,

\begin{eqnarray}
\Delta _{c}^{(p)}(\alpha ,u) &=&{\rm P}\int_{0}^{\infty }\func{Re}\left[ 1-%
\frac{1}{\varepsilon (z,u)}\right] Y_{0}\left( {\cal B}z\right) zdz \\
&&-\frac{2}{\pi }\int_{0}^{\infty }\left[ 1-\frac{1}{\widetilde{\varepsilon }%
(z,u)}\right] K_{0}\left( {\cal B}z\right) zdz.  \nonumber
\end{eqnarray}
The first and second terms in Eq. (36) vanish as ${\cal B}^{-3/2}$ and $%
{\cal B}^{-2}$ respectively at large interionic distances. This leads again
to the LW equipartition rule for point-like projectiles. Thus, in contrast
to the case of an aligned dicluster with $\vartheta =0$ (see Eq. (28)), for $%
\vartheta =\pi /2$ there is no need to introduce an infinitesimal damping
for plasmons and single-particle excitations to get the correct limit at $%
{\cal B}\rightarrow \infty $. This is because a fast ion moving in an
electron gas excites the wake field which has different structures along and
across of the direction of motion \cite{9}. These waves are damped strongly
across the direction of motion \cite{9}.

\subsection{Some simple SPSR}

In this subsection we will consider some useful summation rules. To derive
the first of them we consider the following contour integral

\begin{equation}
Q(\alpha ,u)=\int_{C_{1}}{\cal Z}^{2}(z)\left[ 1-\frac{1}{\varepsilon (z,u)}%
\right] z\ln zdz=-2\pi i\sum_{j}\frac{z_{j}{\cal Z}^{2}(z_{j})}{\frac{%
\partial }{\partial z}\varepsilon (z_{j},u)}\ln z_{j},
\end{equation}
where the contour $C_{1}$ is shown in Fig. 1.

Now, in contrast to the previous sections we take the real part of both
sides of Eq. (37). Then using similar calculational techniques as in Sec.
2.1 we find

\begin{eqnarray}
&&{\rm P}\int_{0}^{\infty }{\cal Z}^{2}(z)\func{Re}\left[ 1-\frac{1}{%
\varepsilon (z,u)}\right] z\ln zdz \\
&=&\frac{\pi }{2}P_{\infty }\left( \alpha ,u\right) -{\rm P}\int_{0}^{\infty
}\widetilde{{\cal Z}}^{2}(z)\left[ 1-\frac{1}{\widetilde{\varepsilon }(z,u)}%
\right] z\ln zdz.  \nonumber
\end{eqnarray}

In a similar way we can obtain some other summation rules

\begin{equation}
{\rm P}\int_{0}^{\infty }{\cal Z}^{2}(z)\func{Re}\left[ 1-\frac{1}{%
\varepsilon (z,u)}\right] zdz=-{\rm P}\int_{0}^{\infty }\widetilde{{\cal Z}}%
^{2}(z)\left[ 1-\frac{1}{\widetilde{\varepsilon }(z,u)}\right] zdz,
\end{equation}

\begin{equation}
{\rm P}\int_{0}^{\infty }\func{Re}\left[ 1-\frac{1}{\varepsilon (z,u)}\right]
z\ln zdz=-\int_{0}^{\infty }\left[ 1-\frac{1}{\widetilde{\varepsilon }(z,u)}%
\right] z\ln zdz  \nonumber
\end{equation}
and

\begin{equation}
{\rm P}\int_{0}^{\infty }\func{Re}\left[ 1-\frac{1}{\varepsilon (z,u)}\right]
zdz=-\int_{0}^{\infty }\left[ 1-\frac{1}{\widetilde{\varepsilon }(z,u)}%
\right] zdz.
\end{equation}
The singularities in the RHS of Eqs. (38) and (39) can be understood in a
similar manner as in Eqs. (34) and (35).

In order to illustrate our analytical results, Eqs.(25)-(28) and (34)-(36),
in Fig. 2 we present $\Delta (\alpha ,u)$ as a function of the parameter $%
u=\omega /kv_{F}$ ($u_{0}\leqslant u\leqslant \lambda $, where $u_{0}$ is
the threshold value for plasmon excitation) for an individual He$^{+}$ ion
(solid line). The lines with and without circles correspond to He$^{+}$ ion (%
$R=3$ \AA ) and diproton ($R=1$ \AA ) clusters, respectively. $\vartheta =0$
(dashed lines) and $\vartheta =\pi /2$ (dotted lines). The numerical results
are for fast projectiles $\lambda =V/v_{F}=8$ and for density parameter $%
r_{s}=2.07$ appropriate to the valence electrons in Al. $\Delta (\alpha ,u)$
as a function of $u$ has interesting features. $\Delta (\alpha ,u)$ remains
positive ($\Im _{sp}>\Im _{p}$) for He$^{+}$ ion and for He$^{+}$ ion
cluster with $\vartheta =0$ (at $u>2$) and $\vartheta =\pi /2$ and remains
negative ($\Im _{sp}<\Im _{p}$) for diproton cluster with $\vartheta =0$ and 
$\vartheta =\pi /2$. The curve for He$^{+}$ ion cluster with $\vartheta =0$
crosses the zero-axis for $u\simeq 2$. From Fig. 2 the oscillatory nature of
the function $\Delta (\alpha ,u)$ can be seen for both diproton and He$^{+}$
ion clusters with $\vartheta =0$. Note that for a diproton cluster with $%
\vartheta =\pi /2$ at $u=\lambda $, when the excited wave moves with the
phase velocity $\omega /k=V$, the parameter ${\cal B}$ vanishes, ${\cal B}=0$%
. Therefore, from Eqs. (36), (40) and (41) at ${\cal B}\rightarrow 0$
follows $\Delta =0$, i.e. the LW equipartition strongly holds. While for He$%
^{+}$ ion cluster with $\vartheta =\pi /2$ from Eqs. (34), (38) and (39) we
find that the function $\Delta $ at $u=\lambda $ is two times greater than
for an individual He$^{+}$ ion projectile. In general for the high velocity
domain $\omega /k\gg v_{F}$ (or $u\gg 1$) all the curves decrease and an
approximate LW equipartition rule, $\Im _{sp}\simeq \Im _{p}$ holds
asymptotically. But it is also clear that in this high velocity limit the
energy losses due to single-particle ($\Im _{sp}$) and plasmon ($\Im _{p}$)
excitations decrease as well.

We may conclude with the remark that our analytical expressions are well
supported by numerical results.\bigskip

\begin{center}
{\bf ACKNOWLEDGMENT\medskip }
\end{center}

Hrachya B. Nersisyan gratefully acknowledges partial financial support by
the International Science and Technology Center (Grant No. A-353).

\hspace{1in}

\newpage

\begin{center}
{\bf Figure Captions}
\end{center}

Fig. 1. Illustration of contours $C_{1}$ and $D_{1}$, in complex $z$ plane.
Isolated point $P$ below real $z$ axis indicates plasmon pole. Group of
crosses above real $z$ axis indicates poles in single-particle excitations.

Fig. 2. The function $\Delta (\alpha ,u)$ vs parameter $u$ for $\lambda =8$.
Individual He$^{+}$ ion (solid line), diproton cluster with $R=1$ \AA , $%
\vartheta =0$ (dashed line without circles) and $\vartheta =\pi /2$ (dotted
line without circles), He$^{+}$ ion cluster with $R=3$ \AA , $\vartheta =0$
(dashed line with circles) and $\vartheta =\pi /2$ (dotted line with
circles). The density parameter is $r_{s}=2.07$ (Al target).

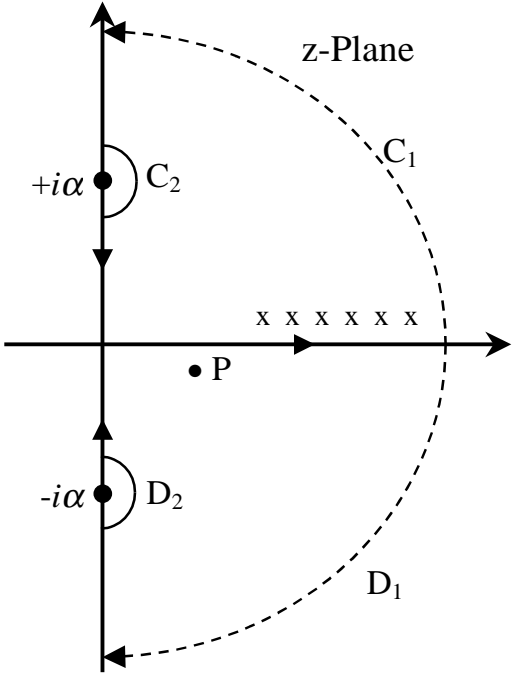
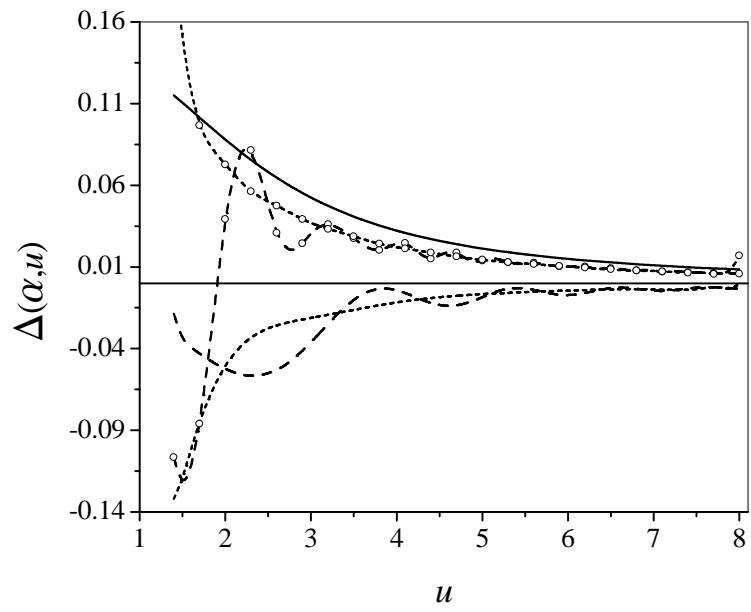
\end{document}